\documentclass[twocolumn,aps,9pt]{revtex4}

\usepackage[dvips]{graphicx}
\usepackage{float,lscape}
\usepackage{tabularx}
\usepackage[sc]{mathpazo}
\usepackage{amsmath}
\begin{document}
\title{Asymmetric adsorption in an open electrolytic cell}
\author{S. Bousiadi, I. Lelidis$\footnote{Corresponding author: ioannis.lelidis@phys.uoa.gr}$}
\affiliation{
 Faculty of Physics, National and Kapoditrian University of Athens,\\
Panepistimiopolis, Zografos, Athens 157 84, Greece}

\date{\today}

\begin{abstract} We investigate the effect of adsorption-desorption phenomenon of ions in an asymmetric electrolytic cell at open circuit conditions. Our approach is based on the Poisson-Nernst-Planck theory for electrolytes and the kinetic model of Langmuir for the description of adsorption-desorption phenomena on the electrodes. When the electrodes are immersed into the solution, selective ion adsorption takes place. It is shown, that the selective ion adsorption is responsible for generating an electrical potential difference between the electrodes of the cell. The analytical expressions for the potential difference and for the charge distribution are calculated. Finally, the time evolution of the system is investigated and the relaxation times of the problem are deduced numerically.
\end{abstract}
% % % % % % % % % % % % % % % % % % % % % % % % % % % % % % % % % % % % % % % %

\maketitle
% % % % % % % % % % % % % % % % % % % % % % % % % % % % % % % % % % % % % % % %

\section{Introduction}

Specific adsorption of ions at interfaces is an omnipresent phenomenon in a large variety of systems such as biological systems, colloids, geological systems, electrolytic cells, liquid crystals, ionic liquids, gels, etc \cite{jrm1,tribolet,bookBE,reddy,kunz,coster}. Specific adsorption of ions (SAdI) starts when an electrode is immersed in a solution or comes into contact with another medium. The effect is determined by ion-electrode, ion-solvent, and solvent-electrode interactions. The SAdI may drive profound effects on both the electrode and the solution with implications for the system itself and for the detection of signals and therefore to all experimental methods that require the use of electrodes, such as Impedance Spectroscopy, Electrocardiogram, Electromyography \cite{biomech}, Impedance Tomography, Bio-Impedance \cite{bioim} etc. In ordered systems such as liquid crystals, adsorption affects molecular orientation, stability of orientation via the surface electric field, anchoring energy, dc-switching of liquid crystal displays, electro-optical modulation, etc \cite{bookBE,lb}.

The adsorption in a electrolytic cell has been mainly considered under an externally applied voltage for electro-optical effects \cite{giov_ad,Thurston1,Thurston2,specificads,compton} and for experimental methods that require the application of an electric field. Recently, the potential difference between an electrode and the bulk of an electrolytic solution has been discussed in absence of an external field \cite{merletti}. We think that it is interesting to consider the potential difference between the electrodes of a cell due to SAdI because it is essential to know the charge distribution in the cell prior to the application of any electric field especially when transient effects are considered.

The goal of the present study is to investigate the establishment of a potential difference and of charge distribution between the electrodes of an electrolytic cell when a dielectric liquid is introduced in the cell, in the absence of any external voltage and for an open circuit. This potential difference should arrive when the two electrodes have different properties related to selective adsorption of ions, that is, the electrodes are asymmetric in what concerns SAdI.
Our approach is developed in the frame of the Poisson-Nernst-Planck (PNP) theory,
that is one of the most used theories describing ion effects in complex systems, and  with many applications in physics, chemistry, and biology \cite{jrm1,tribolet,bookBE,mzb0,mzb1,alai,mzb2,bisquert,luiz,golo,pabst,ross,serghei}. The PNP model is based on the coupled continuity equations of each ion type and the Poisson equation for the electric potential. However, the PNP theory has some well known drawbacks, such like the neglect of finite volume effects and the ion-ion interactions \cite{wang} which are considered in the frame of a mean field approach. Specific adsorption on the electrodes is taken into account by introducing the adequate boundary conditions for the coupled differential equations of the bulk. In our approach we use the Langmuir model \cite{atkins,israel} to describe the kinetics of adsorption-desorption at the electrodes. In order to respect the limitations of the model we assume adsorption is weak, low enough density of ions, and absence of ion generation-recombination effects \cite{angel}.

The rest of the present paper is organised as follows. In section II, we introduce the equations of the PNP model and the adequate BCs according the Langmuir kinetic for adsorption/desorption phenomena. In section III, we linearise the bulk equations and decompose the charge distributions and the potential at a steady state and a transient component. In Section IV, we calculate analytically the charge distribution, the potential in the bulk, and the adsorbed charge at the electrodes. A numerical application is performed for the case of a liquid crystal cell with asymmetric electrodes. In Section V, we treat the transient part of the problem and we show that the evolution towards equilibrium is a multi-relaxation process. The last section of the paper, is devoted to conclusions.

\section{Fundamental equations}

We consider an electrolytic cell composed by an isotropic dielectric liquid between two flat electrodes in parallel position, the left ($L$) electrode located at $z=- d/2$ and the right ($R$) electrode located at $z=+d/2$. The surface area of each electrode is $A_{e\ell}$. The electrolytic solution contains two types of ions denoted $p,m$, where $p$ stands for cations and $m$ for anions, with charge $\pm q$ respectively. In the present analysis, we assume that the density of ions in thermodynamical equilibrium, $N$, is very small with respect to the bulk density of the liquid molecules, so that association-dissociation effects can be neglected \cite{angel}. The ions are supposed point-like and the system as one-dimensional. In the presence of an electric field $E(z,t)$, the current number densities $j_a(z,t)$ are given by

\begin{equation}
j_{a}=-D_{a}\frac{\partial{n_{a}}}{\partial{z}}\pm\mu_an_{a}E
\end{equation}
where $n_a(z,t)$, with $a=p,m$, are the local densities of the ions. $D_a$ are the diffusion coefficients of the
ions which are related with their mobilities via the
Einstein-Smoluchowski relations: $\mu_a/D_a=q/k_BT$ \cite{atkins},
where $k_B$ is the Boltzmann constant and $T$ the temperature.

The fundamental equations that describe the system under investigation are the bulk continuity and Poisson's equations

\begin{eqnarray}\label{continuity}
\frac{\partial{n_{a}}}{\partial{t}} &=&-\frac{\partial{j_{a}}}{\partial{z}}\\
\frac{\partial^2 V}{\partial{z^2}}&=&-\frac{q}{\varepsilon}(n_{p}-n_{m})\nonumber
\end{eqnarray}
where  $\varepsilon$ is the dielectric permittivity of the solvent, and  $V(z,t)$ the electrical potential in the sample.

\subsection{Boundary Conditions}

The boundary conditions of the problem are defined from the current at the electrodes
\begin{eqnarray}
j_a(\pm d/2,t)=\pm\dfrac{\mathrm{d}\sigma_a(\pm d/2,t)}{\mathrm{d}t}
\end{eqnarray}
where $\sigma_a$ are the surface densities of the adsorbed ions. The sign minus applies for the $L$-electrode at $-d/2$ and sign plus for the $R$-electrode at $+d/2$. According to Langmuir kinetic model \cite{lb,giov_ad,atkins}, at the electrodes holds

\begin{equation}\label{Langmuir}
\frac{\mathrm{d}\sigma_{a}}{\mathrm{d}t}=k_{a}n_{a}-\frac{1}{\tau_{a}}\sigma_{a}
\end{equation}
where $k_a$ and $\tau_a$ are the adsorption and desorption coefficients respectively. Langmuir approach holds under the following assumptions: (i) all adsorbing
sites are equivalent, (ii) the surface is uniform, (iii) adsorption occurs only in a first layer, and (iv) the probability of a
particle being adsorbed is independent of the surface density of
particles already adsorbed. These assumptions are valid for low density of ions at thermodynamical equilibrium. Taking into account that $n_{a}\sim N$, whereas the surface density of adsorbed particles has to be $\sigma_{a}<< \frac{1}{l_{a}^2}$,  it follows that $k_{a}\tau_{a}<< \frac{1}{Nl_{a}^2} $, where $l_{a}$ are the typical dimensions of  ions \cite{jrm1}. In the present approach, we consider different adsorption and desorption coefficients for each type of ions and for each  electrode. We indicate by $k_{a}^{i}$, $\tau_{a}^{i}$ the adsorption and desorption coefficients respectively at the electrodes,  $i=L,R$.

Since we consider asymmetric electrodes the two Eqs (\ref{Langmuir}) split to four equations that, in compact form (see appendix-A), are written as

\begin{eqnarray}\label{surfcharge}
\frac{d\sigma^i_a}{dt}=k^i_a n_a^i-\frac{\sigma^i_a}{\tau^i_a}
\end{eqnarray}

Furthermore, the condition on the conservation of the number of ions implies the following equations

\begin{eqnarray}\label{condelect}
\sigma_p^R+\sigma_p^L+\int^{\frac{d}{2}}_{-\frac{d}{2}}n_{p}(z,t)\,\mathrm{d}z=Nd\nonumber\\
\sigma_m^R+\sigma_m^L+\int^{\frac{d}{2}}_{-\frac{d}{2}}n_m(z,t)\,\mathrm{d}z=Nd
\end{eqnarray}

\subsection{Linearisation}

The presence of the electrodes in contact with the electrolyte, creates a small variation of the charge densities $\delta n_{a}<< N$, that become $n_{a}=N+\delta n_{a}$. Henceforth, we introduce the following reduced quantities $\mathcal{P}(z,t)=\frac{\delta n_{p}}{N}$ , $\mathcal{M}(z,t)=\frac{\delta n_{m}}{N}$ , $\mathcal{U}(z,t)=\frac{V}{V_{\theta}}$ where $V_{\theta}=\frac{kT}{q}$ is the thermal voltage. After linearization, the bulk Eqs (\ref{continuity}), are written as

\begin{eqnarray}\label{synexeia}
\frac{\partial \mathcal{A}}{\partial{t}}&=& -D_{a}\left( \mathcal{A}''\pm \mathcal{U}''\right)\\
\label{dynamiko}
\mathcal{U}'' & = & -\frac{\mathcal{P-M}}{2 \lambda^2}
\end{eqnarray}
where the prime means a derivation with respect to $z$, $\mathcal{A=P,M}$, and $\lambda=\sqrt{\frac{\epsilon k_{B}T}{2Nq^2}}$ is the Debye length. Note that when $\mathcal{A}$ stands for cations, $\mathcal{P}$, then $a$ stands also for cations $p$. The bulk densities of currents are given from

\begin{eqnarray}\label{reumata}
j_{a}&=&-ND_{a}(\frac{\partial \mathcal{A}}{\partial z} \pm \frac{\partial \mathcal{U}}{\partial z})
\end{eqnarray}

In terms of the new quantities, conditions (\ref{condelect}) take the form

\begin{eqnarray} %\label{conservationA}
\mathcal{S}^R_p+\mathcal{S}^L_p+\int^{\frac{d}{2}}_{-\frac{d}{2}}\mathcal{P} dz=0\nonumber\\\label{conservationB}
\mathcal{S}^R_m+\mathcal{S}^L_m+\int^{\frac{d}{2}}_{-\frac{d}{2}}\mathcal{M} dz=0
\end{eqnarray}
where $\mathcal{S}_{a}(z,t)=\sigma_{a}/N$.
Eqs (\ref{surfcharge}) that relate the surface charge density to the ions density at the electrodes, are rewritten as

\begin{eqnarray}\label{LangmuirR}
\frac{d\mathcal{S}_a^i}{dt}=k^i_a (1+\mathcal{A}^i)-\frac{\mathcal{S}_a^i}{\tau^i_a}
\end{eqnarray}

At this point we decompose the charge distributions and the potential in their steady state and transient components as follows

\begin{eqnarray}
\mathcal{P}(z,t)&=& p(z)+P(z,t)\nonumber\\\mathcal{M}(z,t) &=& m(z)+M(z,t)\nonumber\\
\mathcal{U}(z,t)&=& u(z)+U(z,t)\label{decomp}\\
\mathcal{S}_p(t)&=& s_p(z)+S_p(z,t)\nonumber\\\mathcal{S}_m(t) &=& s_m(z)+S_m(z,t)\nonumber
\end{eqnarray}
where $p(z)$, $m(z)$, $u(z)$, $s_a(z)$ stand for the steady state part of the quantities $\mathcal{P}(z,t)$, $\mathcal{M}(z,t)$, $\mathcal{U}(z,t)$, and $\mathcal{S}_a(z,t)$  respectively. The corresponding transient components  $P(z,t),M(z,t),U(z,t)$, $S_a(z,t)$ cancel out at long enough times ($t\rightarrow\infty$).
%%%%%%%%%%%%%%%%%%%%%%%%%%%%%%%%%%%%%%%%%%%%%%%%%%%%%%%%%%%%%%%%%%%%%%%%%%%%%%%%%%%%%%%%%%%%

\section{Steady state}

In this Section, we are interested to calculate the potential $u(z)$ in the cell at the equilibrium state. In the steady state, the bulk Eqs (\ref{synexeia},\ref{dynamiko}) simplify at
\begin{eqnarray}
p'+u'=c_1\nonumber\\
m'-u' =c_2\\
 u''  =  -\frac{p-m}{2 \lambda^2}
\end{eqnarray}
where  $c_1,\, c_2 $ are integration constants.
Using the equilibrium condition $\mathrm{d}\sigma/ \mathrm{d}t=0$ or equivalently  $j_p^i=j_m^i=0$, that is, at the electrodes

\begin{eqnarray}
\frac{\partial p}{\partial z}+\frac{\partial u}{\partial z} &=& 0\nonumber\\
\frac{\partial m}{\partial z}-\frac{\partial u}{\partial z} &=& 0
\end{eqnarray}
the continuity equations reduce to $p+u=c_p $ and $ m-u=c_m$ and the Poisson equation is rewritten as
\begin{equation}\label{dynamikoH}
 u''-\frac{u}{\lambda^2}+\frac{C}{\lambda^2}=0
\end{equation}
where $2C=c_p-c_m$. The general solution of the differential equation (\ref{dynamikoH}) is
\begin{equation}
u=A\cosh\frac{z}{\lambda}+B\sinh{\frac{z}{\lambda}}+C
\end{equation}
and the bulk densities of ions are given from
\begin{eqnarray}
p=F-A\cosh(\frac{z}{\lambda})-B\sinh(\frac{z}{\lambda})\\
m=F+A\cosh(\frac{z}{\lambda})+B\sinh(\frac{z}{\lambda})
\end{eqnarray}
with $2F=c_p+c_m$. The integration constants $A,B,F$ are calculated from the conditions on the charge conservation formulated by Eqs (\ref{conservationB}), and the electric field at the electrodes

\begin{equation}\label{E}
\frac{\partial u^i}{\partial z}=\mp\frac{\delta s^i}{2\lambda^2}
\end{equation}
where $\delta s^i=s_p^i-s_m^i$ is the total charge of each electrode ($i=L,R$).
After some algebra we obtain the following expressions for the integration constants in function of $s_a^i$
\begin{eqnarray}\label{A}
A&=&\dfrac{s^R_p+s^L_p-s^R_m-s^L_m}{4\lambda\,\sinh\left(\dfrac{d}{2\lambda} \right) }\\\label{B}
B&=&\dfrac{s^R_p-s^R_m - (s^L_p-s^L_m)}{4\lambda\,\cosh\left(\dfrac{d}{2\lambda} \right) }\\\label{F}
F&=&-\dfrac{s^R_p+s^L_p+s^R_m+s^L_m}{2d}
\end{eqnarray}

and Eqs (\ref{LangmuirR}) using the condition $\mathrm{d} \sigma_a^i/\mathrm{d}t=0$, yield

\begin{equation}\label{s}
s_a^i=k_a^i\tau_a^i\,(a^i+1)
\end{equation}
The analytical expressions calculated for the adsorption-desorption terms $s_a^i$ are given in the appendix-A.

Finally, the potential difference between the two electrodes due to selective adorption phenomena is given  from the equation
\begin{eqnarray}\label{Du}
\Delta u=u^R-u^L=\dfrac{\delta s^R-\delta s^L}{4\lambda}\,\tanh\left( \frac{d}{2\lambda}\right)
\end{eqnarray}
 For a symmetric cell  $\Delta u=0$. For an asymmetric cell $\Delta u\ne 0$ in general. For electrodes with preference to opposite sign of charge and for $d>>\lambda$, $\Delta u$ becomes maximal, up to a value of $2$, when the Debye length  $\lambda<<k\tau$ the sorption mechanism length. It is well known that larger values of the potential than the thermal voltage can be present at an electrode \cite{jap} but in this latter case our linear approximation is no longer valid and one should consider the non-linear problem.

 % % % % % % % % % % % %FIG1 % % % % % % %
% % % % % % % % % % % % % % % % % % %

\begin{figure}
\includegraphics[width=5.1cm]{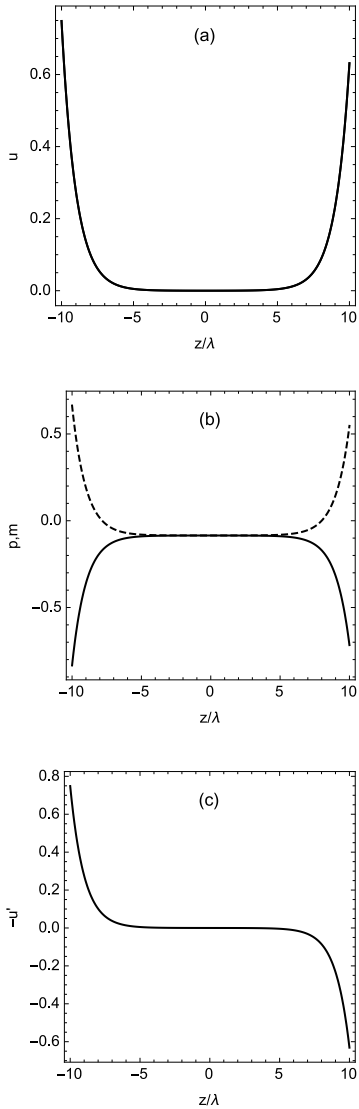}
\caption{ Asymmetric cell, the effective adsorption is dominated from charge of the same sign on both electrodes. (a) potential $u(z)$ at the steady state, (b) distribution of the ions $p(z)$ (solid line), $m(z)$ (dashed line), and (c) electric field $-u'(z)$. The Langmuir model parameters used for the calculation are $K_p^R=5$,  $K_m^R=0.1$, $K_p^L=10$, and $K_m^L=0.1$.}
\label{fig1}
\end{figure}

% % % % % % % % % % % % %FIG2 % % % % % % %
% % % % % % % % % % % % % % % % % % %

\begin{figure}
\includegraphics[width=6cm]{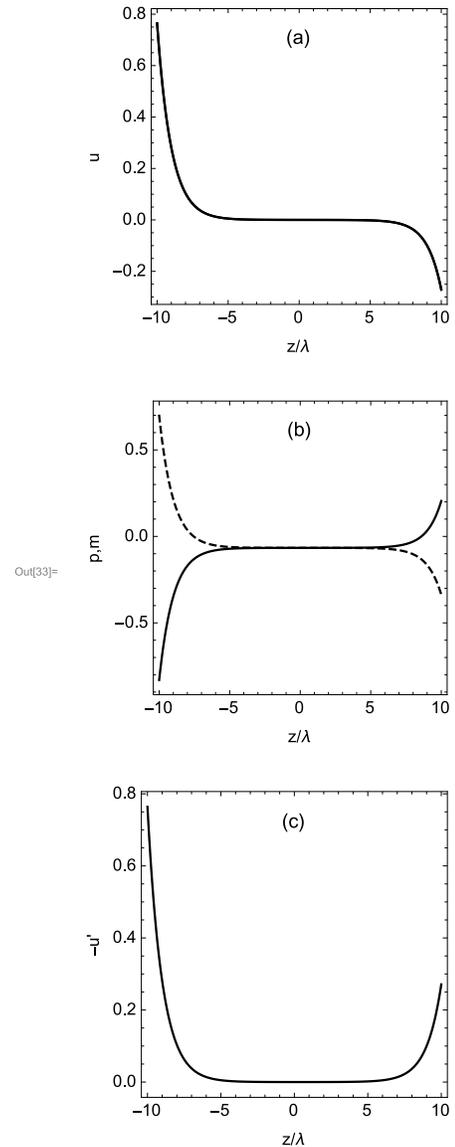}
\caption{ Asymmetric cell, the effective adsorption is dominated from charge of the opposite sign on both electrodes. (a) potential $u(z)$ at the steady state, (b) distribution of the ions $p(z)$ (solid line), $m(z)$ (dashed line), and (c) electric field $-u'(z)$. The Langmuir model parameters used for the calculation are $K_p^R=0.1$,  $K_m^R=1$, $K_p^L=10$, and $K_m^L=0.1$.}
\label{fig2}
\end{figure}
\subsection*{Numerical analysis}

Let us consider the case of a symmetrical $1:1$ electrolyte with monovalent ions, and assume $N=10^{22}\mathrm{m}^{-3}$, $d=10^{-5}\mathrm{m}$, $A=10^{-4}\mathrm{m}^2$, $V_{\theta}=0.026 V$. For a liquid with a relative effective \cite{effepsilon} dielectric constant $\varepsilon =10$, the Debye length is $\lambda=2.7\times 10^{-8}$m$<<d$.

\begin{figure}
\includegraphics[width=6cm]{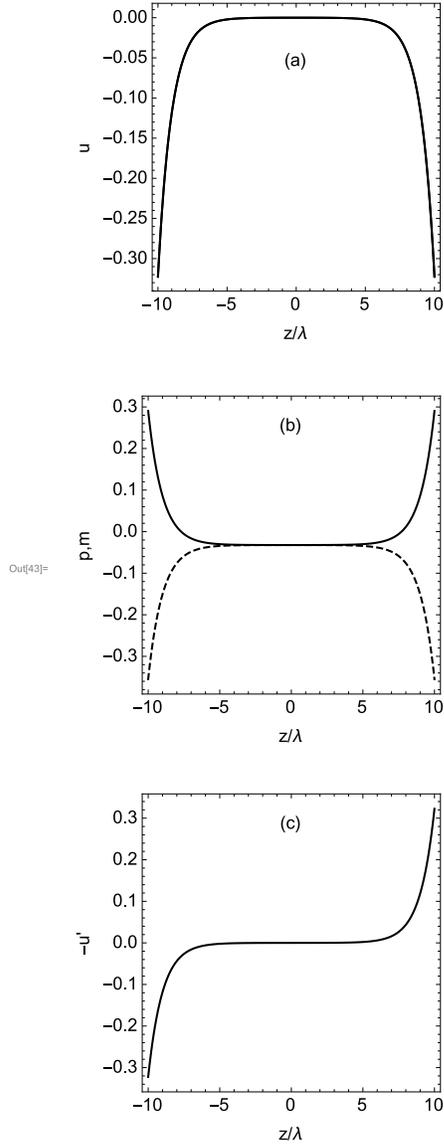}
\caption{Symmetric cell, (a) potential $u(z)$, (b) distribution of the ions $p(z)$ (solid line), $m(z)$ (dashed line), and (c) electric field $-u'(z)$, at the steady state. $K_m^R=K_m^L=1$, and $K_p^R=K_p^L=0$.}
\label{fig3}
\end{figure}
\begin{figure}
\includegraphics[width=6cm]{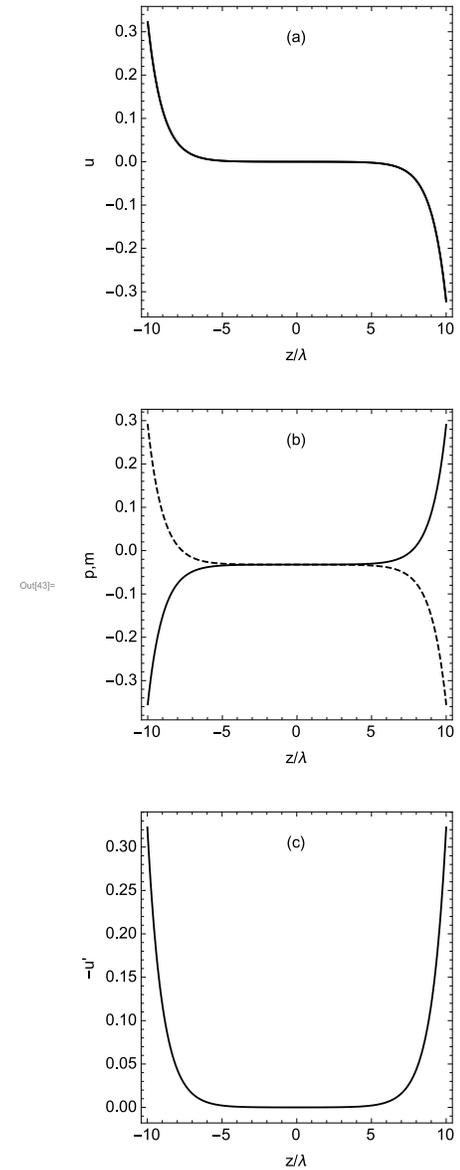}
\caption{Antisymmetric cell,  (a) potential $u(z)$, (b) distribution of the ions $p(z)$ (solid line), $m(z)$ (dashed line), and (c) electric field $-u'(z)$, at the steady state. $K_m^R=K_p^L=1$, and $K_p^R=K_m^L=0$.}
\label{fig4}
\end{figure}

Figure 1, shows in reduced units (a) the potential $u(z)$ as function of the position in the cell at the steady state, (b) the distribution of the ions $p(z)$ (solid line), $m(z)$ (dashed line), and (c) the electric field $-u'(z)$ for an asymmetric cell with effective adsorption dominated from charge of the same sign on both electrodes.  The Langmuir model parameters used for the calculation are $K_p^R=0.1$,  $K_m^R=1$, $K_p^L=0.1$, and $K_m^L=10$, where $K_a^i=k_a^i\tau_a^i$ (see appendix-A). Typical values of the adsorption parameters for liquid crystals are given in \cite{maximus0,maximus,huang,hama}.  Figure 2, shows in reduced units (a) $u(z)$, (b) $p(z)$ (solid line), $m(z)$ (dashed line), and (c) $-u'(z)$ for asymmetric electrodes that are dominated by adsorption of ions that have an effective charge of opposite sign at the electrodes. From both figures is deduced that the potential in the cell is asymmetric. The potential difference between the two electrodes can in general attain values of the same order of magnitude as the thermal voltage (see Figure 2). This observation shows that for asymmetric adsorption the induced potential can have a non negligible effect on impedance spectroscopy measurements for the latter experimental method makes use of a measuring external voltage that in principle should be less than the thermal voltage \cite{bal}.

In the case of a symmetric cell, Figure 3, the electrodes are  identical, that means same adsorption and desorption coefficients for both electrodes, and the problem is reduced to the one already investigated in \cite{merletti}. In the antisymmetric case, Figure 4, the adsorption and desorption coefficients are the same for opposite sign charges at both electrodes. Note that with increasing ratio $d/\lambda>> 1$ the surface values of the parameters do not change (see for instance Eqn(\ref{Du})), therefore we used $d/\lambda=20$ for the plots in order to better visualise the variation of voltage, electric field and charge density near the electrodes.

%%%%%%%%%%%%%%%%%%%%%%%%%%%%%%%%%%%%%%%%%%%%%%%%%%%%%%%%%%%%%%%%%%%%%%%%%%%%%%%%%%%%%%%%%%%%%%%%%%%

\section{Evolution of the system towards the equilibrium state}

%%%%%%%%%%%%%%%%%%%%%%%%%%%%%%%%%%%%%%%%%%%%%%%%%%%%%%%%%%%%%%%%%%%%%%%%%%%%%%%%%%%%%%%%%%%%%%%%%%%
In the present section, we investigate the evolution of the system towards the equilibrium state, that is, we consider the evolution of the transient components  $A(z,t),U(z,t)$, and $S_a(z,t)$. The quantities $P(z,t),M(z,t),U(z,t)$ are derived from the bulk Eqs (\ref{synexeia})

\begin{eqnarray}\label{bulktr}
\frac{\partial P}{\partial t}&=& D_{p}(P'' + U'')\nonumber\\
\frac{\partial M}{\partial t}&=& D_{m}(M'' - U'')
\\
U''&=& -\frac{P-M}{2 \lambda^2}\nonumber
\end{eqnarray}
The initial conditions of the problem are deduced from the fact that before the adsorption-desorption phenomenon takes place, at $t=0$,  $n_a(z,0)=N$ that implies $A(z,0)+a(z)=0$. The same is valid for the rest of Eqs (\ref{decomp}).

The boundary conditions at the electrodes are \begin{eqnarray}\label{BCt}
-D_a(A'\pm U')^i=\pm\dfrac{\mathrm{d}S_a^i}{\mathrm{d}t}
\end{eqnarray}
where the plus sign at the \textit{rhs} applies for the $R$-electrode.
The Langmuir equation for each electrode is written

\begin{equation}\label{LangmuirF}
\frac{dS_{a}^i}{dt}=k^i_a P^i-\frac{1}{\tau^i_a}S_a^i
\end{equation}

To solve the system of differential Eqs(\ref{bulktr}), we eliminate $U''$ from the first two equations via the third one, and use trial solutions of the type:

\begin{equation}\label{trial1}
X(z,t)=y_{x}(z)\exp(-\beta t)
\end{equation}

where $\beta>0$ in order to satisfy the condition $\lim\limits_{t \rightarrow \infty}X=0$, $X=A,U$ and $x=a,u$.

After substitution of the trial solutions (\ref{trial1}) into the bulk Eqs(\ref{bulktr}) and elimination of the Poisson equation, we find the following system of equations

\begin{eqnarray}\label{coupled}
y''_{p}+\gamma_p y_{p}+\frac{y_m}{2\lambda^2}=0\nonumber\\
y''_{m}+\gamma_m y_{m}+\frac{y_p}{2\lambda^2}=0
\end{eqnarray}
with:\begin{equation}
\gamma_a=\frac{\beta}{D_a}-\frac{1}{2\lambda^2}
\end{equation}
To solve the system of Eqs(\ref{coupled}), we seek for solutions of the type
\begin{equation}\label{trial2}
y_a=C_a\,\mathrm{e}^{\mu z}
\end{equation}
After substitution of the latter ansatz in the Eqs(\ref{coupled}), one finds
\begin{equation}\label{quartic}
\begin{split}
&C_p (\gamma_p+\mu^2)+\frac{1}{2\lambda^2}C_m=0\\
&C_m (\gamma_m+\mu^2)+\frac{1}{2\lambda^2}C_p=0
\end{split}
\end{equation}
The solution of the characteristic quartic equation, resulting from the above system, gives for $\mu$ the four solutions: $\pm \mu_1$, $\pm \mu_2$ (their explicit expressions are given in appendix-A). The solution of Eqs(\ref{coupled}) are then written
\begin{widetext}
\begin{eqnarray}\label{y_x}
y_p &=& C_1\mathrm{e}^{\mu_1 z}+C_2\mathrm{e}^{-\mu_1 z}+C_3\mathrm{e}^{\mu_2 z}+C_4\mathrm{e}^{-\mu_2 z}\nonumber\\
y_m &=& k_1C_1\mathrm{e}^{\mu_1 z}+k_1C_2\mathrm{e}^{-\mu_1 z}+k_2C_3\mathrm{e}^{\mu_2 z}+k_2C_4\mathrm{e}^{-\mu_2 z}\\
y_u &=& C_0+C_{\ell}z-\dfrac{1-k_1}{2\lambda^2\mu_1^2}\left(C_1\mathrm{e}^{\mu_1 z}+C_2\mathrm{e}^{-\mu_1 z}\right) -\dfrac{1-k_2}{2\lambda^2\mu_1^2}\left(C_3\mathrm{e}^{\mu_2 z}+C_4\mathrm{e}^{-\mu_2 z}\right) \nonumber
\end{eqnarray}
\end{widetext}

The Langmuir Eqs (\ref{LangmuirF}), using the trial solution Eq (\ref{trial1}), yield

\begin{equation}
S_a^i(t)= \zeta_a^i\mathrm{e}^{-t/\tau_p^i} +K_a^i\,y_a^i\mathrm{e}^{-\beta t}
\end{equation}
where $K_a^i=\frac{k_p^A \tau_p^A }{1-\beta \tau_p^A}$. By substituting the above equations into the BCs given from Eqs(\ref{BCt}), one finds $\zeta_a^i=0$, and the system of linear differential equations
\begin{equation}\label{BCt2}
D_a(y'_a\pm y'_u)^i=\pm\beta K_a^i\,y_a^i
\end{equation}
where in the \textit{lhs} the $+$ applies for $a=p\,\&\,A=P$, and in the  \textit{rhs} the $+$ applies for the $i=R$-electrode. Using the transient part of Eqs(\ref{E}) one can calculate $C_{\ell}$. Substitution of $y_x$, given from Eqs(\ref{y_x}), into Eqs(\ref{BCt2}) results to the homogeneous system of equations

\begin{equation}
\begin{split}
&\alpha_{11}C_1+\alpha_{12}C_2+\alpha_{13}C_3+\alpha_{14}C_4=0\\
&\alpha_{21}C_1+\alpha_{22}C_2+\alpha_{23}C_3+\alpha_{24}C_4=0\\
&\alpha_{31}C_1+\alpha_{32}C_2+\alpha_{33}C_3+\alpha_{34}C_4=0\\
&\alpha_{41}C_1+\alpha_{42}C_2+\alpha_{43}C_3+\alpha_{44}C_4=0
\end{split}
\end{equation}
where the coefficient $\alpha_{nj}$ are functions of the relaxation time $1/\beta$. This eigenvalue problem has a non trivial solution, if and only if, its determinant of coefficients, $\Delta$, equals zero. This condition determines all the characteristic relaxation times $1/\beta_{\ell}$  of the system. Finally, because of the linearity, one can apply the superposition principle to write the complete solution of the problem in the form

\begin{eqnarray}
X(z,t)=\sum_{\left\lbrace \beta_{\ell}\right\rbrace }\,y_x\left(\beta_{\ell},z \right) \,\mathrm{e}^{-\beta_{\ell}t}
\end{eqnarray}
Plotting the determinant $\Delta$,  see Figure 5, one observes  that the system has an infinite number of characteristic time constants corresponding to the roots of the equation $\Delta(\beta_{\ell}) =0$.  Only a few of them should have a physical meaning \cite{marco,marco2,kania}. The overall relaxation effect should be dominated from the slowest relaxation time that is expected to correspond at the adsorption-desorption dynamics at the electrodes. These multiple relaxation time constants are expected, as relaxation depends on a variety of phenomena such as diffusion, adsorption, desorption, and couplings of different mechanism, for instance, the ambipolar diffusion effect \cite{ambi1,ambi2}.

\begin{figure}
\includegraphics[width=8cm]{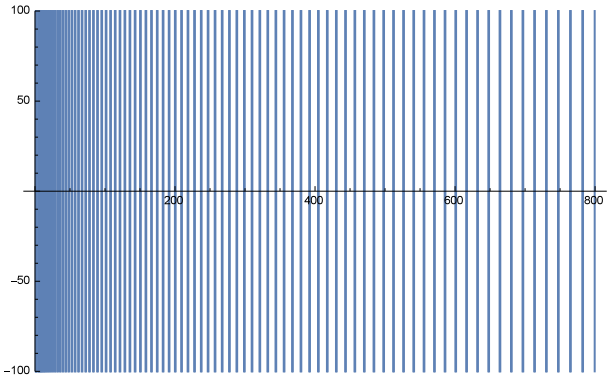}
\caption{The determinant $\Delta$ vs $\beta$ in arbitrary units. The zeros of $\Delta$ correspond to the inverse relaxation time constants $\beta_{\ell}$ of the system.}
\label{fig5}
\end{figure}
Some of the expected relaxation times that should contribute to the multirelaxation phenomenon are discussed in brief here. We distinguish them to those originating from the  bulk of the electrolyte and those related with the adsorption mechanism.  Since, the cations and ions have different diffusion constants, one can define the ambipolar $D_{amb}=\frac{2D_pD_m}{D_p+D_m}$ and the free diffusion $D_f=\frac{D_p+D_m}{2}$ constants \cite{ambi1,ambi2}. Consequently a few of the bulk diffusion times that are expected should be  $\tau_{amp}\sim d^2/D_{amb}$, $\tau_f\sim \lambda^2/D_{f}$, $\tau\sim d\lambda/D_{amb}$, $\tau\sim d\lambda/D_{f}$, $\tau\sim \tau_f\sqrt{d/\lambda}$, etc.
Since now, in our approach we considered different adsorption and desorption coefficient for each type of ions and for each electrode, we expect eight relaxation times related to the adsorption--desorption phenomenon.
Finally, in order to estimate the influence of the above relaxation times on the overall response of the system one should perform a numerical analysis by calculating the temporal evolution of the system for different values of the model parameters, and try to fit the numerical data as in \cite{marco,marco2}.

\section{conclusion}

We have investigated the phenomenon of selective adsorption in an asymmetric electrolytic cell, that is, the adsorption and desorption coefficients are different for each sign of ions, and for each electrode. The ions have different mobility. We derived the analytic expression for the potential and the charge distribution in the cell, in the frame of the PNP model and in the linear approximation. Our linear numerical analysis has shown that the asymmetry of the electrodes leads to a difference of potential between the electrodes that can attain values of the same order of magnitude as the thermal voltage.
Finally, we have investigated the evolution of the system towards the equilibrium state, always in the case of asymmetric adsorption. The system has infinite characteristic times, but only a few of them are expected to have a physical interest.

%\section*{ACKNOWLEDGMENTS}

% % % % % % % % % % % % % % % % % % % %

% % % % % % % % % % % % % % % % % % % %

%\newpage

% % % % % % % % % % % % % % % % % % % %
\appendix
\section{}
Langmuir kinetic model equations at the electrodes (Eqs (\ref{surfcharge})):
\begin{eqnarray}
\frac{d\sigma^A_p}{dt}=k^{A}_{p}n_{p}-\frac{\sigma^A_p}{\tau^{A}_{p}}\\
\frac{d\sigma^A_m}{dt}=k^{A}_{m}n_{m}-\frac{\sigma^A_m}{\tau^{A}_{m}}\\
\frac{d\sigma^B_p}{dt}=k^{B}_{p}n_{p}-\frac{\sigma^B_p}{\tau^{B}_{p}}\\
\frac{d\sigma^B_m}{dt}=k^{B}_{m}n_{m}-\frac{\sigma^B_m}{\tau^{A}_{m}}
\end{eqnarray}

The solution of the equation system  \eqref{A}--\eqref{s} yields the steady state part of the adsorbed ions surface densities at the electrodes:
\begin{widetext}
\begin{eqnarray}
s^R_p &=& -\dfrac{4dK_p^R\sinh^2\delta}{\mathcal{D} }
\left[( K_m^R + K_p^L)\lambda+ K_m^R(K_m^L + K_p^L)\coth\delta  +2\lambda^2\coth\delta +( K_m^R +  K_m^L) \lambda \coth^2\delta\right] \\
 s^R_m &=& -\dfrac{4dK_m^R\sinh^2\delta}{\mathcal{D} }
\left[( K_m^L +  K_p^R)\lambda+ K_p^R(K_m^L + K_p^L)\coth\delta  +2\lambda^2\coth\delta +( K_p^R +  K_p^L) \lambda \coth^2\delta\right] \\
s^L_p &=& \dfrac{2dK_p^L}{\mathcal{D} }
\left[\lambda( K_m^R -  K_p^R)+ \lambda(2K_m^L + K_m^R+K_p^R)\cosh 2\delta  +( K_m^LK_m^R +  K_m^LK_p^R+2\lambda^2)  \sinh 2\delta\right] \\
s^L_m &=& \dfrac{2dK_m^L}{\mathcal{D} } \left[\lambda(  K_p^R-K_m^R)+ \lambda(2K_p^L + K_m^R+K_p^R)\cosh 2\delta  +( K_p^LK_m^R +  K_p^LK_p^R+2\lambda^2)  \sinh 2\delta\right]
\end{eqnarray}
where
\begin{eqnarray*}
\mathcal{D} &=& -2\lambda(K_m^L-K_p^L)(K_m^R-K_p^R)+2\lambda  \left(\varPi_2 +\varSigma_K d\right)\cosh 2\delta
+ \left[ d(K_p^R+K_m^R)(K_p^L+K_m^L)+4\lambda^2d+2(\varPi_3 +\varSigma\lambda^2)\right] \sinh 2\delta\\
\\
\delta &=& \dfrac{d}{2\lambda}\\
K_a^i &=& k_a^i\tau_a^i\\
\delta s^i&=& s_p^i-s_m^i\\
\varSigma &=& K_p^R+K_m^{R}+K_p^{L}+K_m^L\\
\varPi_2 &=& K_p^R K_p^L +  K_m^L K_m^R+K_p^R K_m^L+K_p^L K_m^R+ 2 K_p^R  K_m^R+ 2 K_p^L  K_m^L\\
\varPi_3 &=& K_p^R K_p^L K_m^R +K_p^R  K_p^L K_m^L+ K_p^R K_m^L K_m^R+ K_p^L K_m^L K_m^R\\
\varPi_4 &=& K_p^R K_m^{R} K_p^{L} K_m^L\\
\end{eqnarray*}
\end{widetext}
Solutions of the quartic equation resulting from the equation system (\ref{quartic}):
\begin{eqnarray}
\mu_1=\sqrt{-\dfrac{\gamma_p+\gamma_m}{2}-\dfrac{\sqrt{1+\lambda^4(\gamma_p-\gamma_m)^2}}{2\lambda^2}}\\
\mu_2=\sqrt{-\dfrac{\gamma_p+\gamma_m}{2}+\dfrac{\sqrt{1+\lambda^4(\gamma_p-\gamma_m)^2}}{2\lambda^2}}
\end{eqnarray}
Integration constants of the equation system (\ref{y_x}):
\begin{equation}
k_1=-2 \lambda^2 \left(\gamma_p+\mu_1^2\right)
\end{equation}

\begin{equation}
k_2=-2 \lambda^2 \left(\gamma_p+\mu_2^2\right)
\end{equation}

\end{document}